\documentclass[%
reprint,
 amsmath,amssymb,
 aps,
]{revtex4-1}

\usepackage{graphicx}
\usepackage{subfigure}
\usepackage{dcolumn}
\usepackage{bm}

\begin{document}

\title{A New Strategy in Applying the Learning Machine to Study Phase Transitions}

\author{Rongxing Xu}
\author{Weicheng Fu}
\author{Hong Zhao}%
 \email{zhaoh@xmu.edu.cn}
\affiliation{%
 Department of Physics and Jiujiang Research Institute,
 Xiamen University, Xiamen 361005, Fujian, People's Republic of China
}%

\date{\today}

\begin{abstract}
In this Letter, we present a new strategy for applying the learning machine to study phase transitions. We train the learning machine with samples only obtained at a non-critical parameter point, aiming to establish intrinsic correlations between the learning machine and the target system. Then, we find that the accuracy of the learning machine, which is the most important performance index in conventional learning machines, is no longer a key goal of the training in our approach. Instead, relatively low accuracy of identifying unlabeled data category can help to determine the critical
point with greater precision, manifesting the singularity around the critical point. It thus provides a robust tool to study the phase transition. The classical ferromagnetic and percolation phase transitions are employed as illustrative examples.

\end{abstract}

\maketitle

\emph{Introduction --- }Machine learning has become an new tool for solving various physical problems \cite{girshick2014rich,Toshev2014DeepPose,6909619,Jordan2015Machine,
PhysRevB.96.195145,PhysRevB.97.054303,PhysRevB.97.045153,PhysRevLett.120.257204,
PhysRevA.97.042315,PhysRevLett.121.150503,PhysRevE.98.053305,PhysRevB.96.205146,
PhysRevLett.120.066401} such as crystal structure prediction \cite{PhysRevLett.91.135503,PhysRevB.97.054303,PhysRevE.98.053305}, quantum problems \cite{carleo2017solving,PhysRevB.90.155136,PhysRevLett.120.240501,PhysRevLett.118.216401,
PhysRevA.97.042315,PhysRevLett.121.150503,PhysRevLett.120.066401} and, in particular, the identification of phase transitions \cite{PhysRevB.94.165134,PhysRevLett.118.216401,carrasquilla2017machine,
van2017learning,portman2017sampling,PhysRevB.94.195105,PhysRevLett.118.216401,
PhysRevLett.120.257204,PhysRevB.96.205146}. The fundamental benefit of applying machine learning in physical problems is that this approach can extend beyond the limits of conventional physical approaches by obtaining solutions based on partial or even no prior physical knowledge, and thereby extrapolate them to unexplored data. This benefit has been demonstrated in phase transition problems by unsupervised learning approaches \cite{PhysRevB.94.165134, van2017learning, PhysRevB.94.195105,PhysRevE.98.053305}. Wang \cite{PhysRevB.94.195105} first applied principle component analysis (PCA) to classify the two phases in the Ising model. Later, van Nieuwenburg \emph{et al}. \cite{van2017learning} proposed the so-called confusion scheme to obtain critical points successfully for several Ising-like models. The application of unsupervised machine learning represents a landmark in the study of phase transitions.

However, these researches are based on Ising-like models of which configurations are the results of dynamic evolutions. Here, dynamic evolution drives the system to be different clustered configurations that are distinguishable according to the statistical characteristics of the configuration vectors. For other models, phase features are not always rigidly correlated to statistical properties. For example, configurations of the two-dimensional (2D) square-lattice site percolation model are represented by a binary vector with open $(+1)$ and closed $(-1)$ components. The open or closed state of each site is randomly determined according to the probability $p$ or $(1-p)$, where $p$ is the probability of the open state for a site. A percolation configuration is defined as a specific route that is connected from one side of the square to the opposite by sites residing in open states \cite{broadbent1957percolation}, which is not a consequence of dynamic evolution. This is similar to defining the shape of a smile in the binary image, which is a feature with no simple correlation to the intrinsic properties of the configuration vectors. In the case of percolation models then, the information pertaining to a desired phase feature must be included by supervisors.

On the other hand, supervised learning approaches have already been applied to phase transition problems \cite{PhysRevLett.118.216401, carrasquilla2017machine,PhysRevB.96.205146,PhysRevLett.120.257204,PhysRevB.97.045207,
PhysRevLett.120.066401}. Carrasquilla \emph{et al}. \cite{carrasquilla2017machine} demonstrated that a well-designed learning machine can distinguish phases of some Ising-like models. Nevertheless, labeled samples obtained over the entire parameter interval were required in the training process. Then, phase transition curves can be obtained by simply counting the number of labeled samples, and the learning machine merely plays a role to enhance the accuracy of the results by identifying a large amount of unlabeled configurations. Other researches based on supervised learning have the similar situations  \cite{PhysRevB.96.205146,PhysRevLett.120.257204,PhysRevB.97.045207,PhysRevLett.120.066401}. Even though some research declared that they can successfully train a learning machine by the configurations obtained near the critical point \cite{PhysRevLett.120.240501}, it implicitly applies the prior-information of critical point, which should be unknown here before the learning machine studies. Therefore, traditional supervised learning cannot go beyond the physical approaches to find the critical points in phase transition problems and a new strategy is needed.

In this Letter, we first illustrate that the unsupervised methods fail in identifying phases in percolation models, and then clarify the conditions under which these methods work. Next, we introduce the new strategy by using the 2D square-lattice site percolation model and the 2D Ising model as examples. On the framework of machine learning, it appears as a semi-supervised learning. We generate configurations at an arbitrary parameter point to train the learning machine, and apply the machine to identify unlabeled data over the entire parameter interval. Different from the conventional supervised learning, the accuracy is not the central factor. In more detail, in usual application of learning machine, the accuracy of identifying data categories is the most important performance index, and the higher the better. Here, instead of pursuing a high accuracy, the purpose of training is to  establish an intrinsic correlation between the learning machine and the target system, leading a singularity around the critical point. The singularity may be manifested as the maximum or minimum uncertainty in identifying data categories, or sudden changes in some observable quantities. Therefore, the unorthodox application of the learning machine provides a robust tool to recover phase diagram, and even to find sub-phase structures that conventional physical approaches may ignored.

\begin{figure}[h]
  \centering
  \includegraphics[width=1\columnwidth]{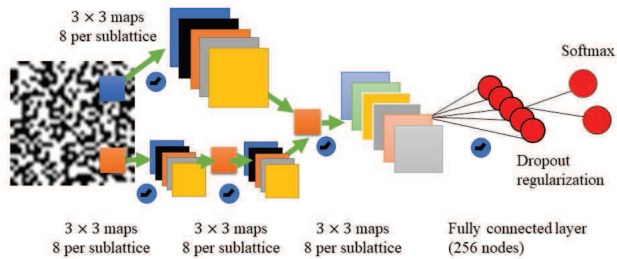}
  \caption{Architecture of the convolutional neural network.}\label{1}
\end{figure}

\emph{Learning Machine and Models --- }
The machine learning proposed in the present work employs a convolutional neural network (CNN). The architecture is illustrated in Fig. 1. The convolutional layer is divided into two parts, which include a layer with eight $3 \times 3$ filters and a small three-layer network with eight $3 \times 3$ filters per layer. A rectified linear unit is adopted as the activation function. The output layer adopts a classical softmax classifier following a fully connected layer with 256 rectified linear units \cite{szegedy2016rethinking,SupplementMaterial}.

Two models, the 2D square-lattice site percolation model and the 2D Ising model, are applied in our study. The former one has been already introduced above. The latter is defined on the square lattice with the nearest-four spin-spin interactions by the Hamiltonian, $H = -\sum_{ij} \sigma_i \sigma_j - h\sum_i^N \sigma_i$, where the spin $\sigma = \pm 1$. Periodic boundary conditions are adopted here. In the thermodynamic limit with $h = 0$, a second-order phase transition occurs at the critical temperature $T_c$ ($\approx 2.269$), beyond which configurations will simultaneously converge to the non-ferromagnetic phase for which $\langle m \rangle = 0$, where $\langle m \rangle$ is the magnetization defined as $\langle m \rangle = \frac{1}{N}\sum_{i=1}^N \sigma_i$ \cite{Newman1999Monte}. Mathematically, configurations for both models can all be regarded as square binary images. For machine learning each image is converted to a 1D vector according to orders of its rows. Besides, the exact categories of configurations are determined by the two-pass algorithm \cite{shapiro2001computer} for the percolation model and the traditional Metropolis Monte-Carlo methods \cite{Newman1999Monte} for the Ising model.

\emph{Unsupervised Learning --- }We first illustrate that the unsupervised learning methods, including K-means algorithm, the confusion scheme and the PCA method, fail in finding the percolation phase transition. Here, the training set is obtained by sampling $10^3$ configurations every $0.01$ at the entire parameter interval of $p$. K-means algorithm aims to figure out the training set into $k$ clusters in which each data point belongs to the cluster with the nearest mean \cite{macqueen1967some}. To cluster configurations into two classes, we initialize $2$ clustering centroids randomly, calculate the Euclidian distances from each configuration to every centroid, find the nearest one to the configuration and tag it with the label of the centroid. Positions of clustering centroids can be obtained by updating the averaging positions of all the  points of configurations until they converge. With these centroids, labels can be tagged and the proportion $P$ of percolating configurations at every $p$ in the total set can be calculated. Fig. 2 (a) shows the predicted phase transition curve for two choice of the data set. One is from the interval $p \in [0.3, 1]$ and the other from $p \in [0, 1]$, respectively. It can be found that the predicting critical point is not in accordance with the real one ($0.593$). Instead, the predicted curves depend on the intervals of data sets.

To apply the confusion scheme, we need to guess a sequence of critical point $c'$ in the parameter interval $[c_1, c_2]$ and assign configurations at the two sides divided by $c'$ with different labels respectively. Thus, a sequence of training and testing data sets are constructed. Then, a learning machine trained by these training sets with different boundary points $c'$ should provide a sequence of accuracies on the corresponding testing set and show a "W-shape" in the range $[c_1, c_2]$ of $p$, of which the middle peak is expected to be corresponding to the critical point \cite{van2017learning}. However, in Fig. 2 (b) we can see a "V-shape" performance curve with a minimum value at approximately $p = 0.5$. Therefore, the confusion scheme also fails in the percolation model.

\begin{figure}[t]
  \centering
  \includegraphics[width=0.5\textwidth]{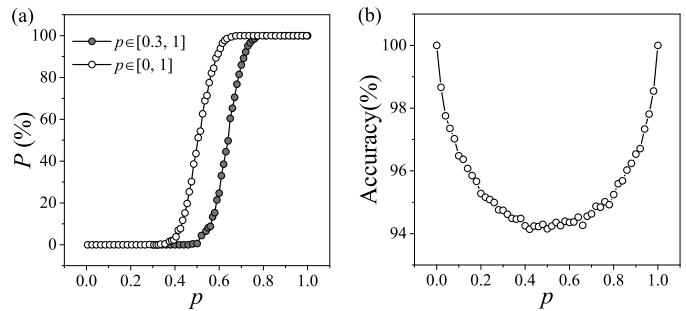}\\
  \caption{Application of unsupervised learning to the percolation model. (a) Phase-transition curves obtained by the k-means clustering algorithm indicating the predicted percentage $P$ of percolating configurations in the parameter interval $p \in [0,1]$ or $p \in [0.3,1]$. (b) The accuracy of the confusion scheme. The solid lines in the figures are provided merely for guiding the eyes.}\label{2}
\end{figure}

\begin{figure}
  \centering
  \includegraphics[width=0.5\textwidth]{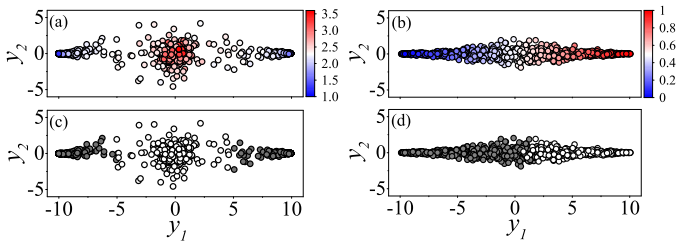}\\
  \caption{PCA results for the Ising model (a) and (c), and the percolation model (b) and (d). The color scales in (a) and (b) denote the temperature $T$ and the open probability $p$, respectively. Filled and open circles in (c) denote the ferromagnetic and non-ferromagnetic phases of the Ising model, respectively, while those in (d) denote the percolating and non-percolating configurations, respectively.}\label{3}
\end{figure}

Last, we check PCA. For the purpose of comparison, we investigate both models here. Let $N$ be the dimension of the vector of a configuration, such that $n$ configurations construct a $n \times N$ sample matrix $X$. In PCA, the equation $X^TX \omega_i = \lambda_i \omega_i$ is solved to obtain $N$ eigenvalues and eigenvectors. The corresponding eigenvectors of the first several eigenvalues are considered as principle components, according to the fact that larger eigenvalues carry more information. By projecting the original sample matrix to these principle components, the feature space is significantly reduced. As such, PCA represents a linear transformation that can extract mutually orthogonal directions along which samples are distributed according to their intrinsic correlations, and can therefore provide basic information for clustering samples into different classes \cite{PhysRevB.94.195105}. In particular models studied here, only the first eigenvector is the dominant component \cite{SupplementMaterial}. For a better visual effect, we show the PCA results by 2D plots by the first two components in Fig. 3. For the Ising model, points of configurations are manifested into three clusters, corresponding to two low-temperature clusters and one high-temperature cluster in Fig. 3 (a), which are respectively coincident with the ferromagnetic and non-ferromagnetic phases in Fig. 3 (c). The clustering structure then can be easily extracted by the K-means algorithm. Also, the confusion scheme works in this situation since at $T_c$ there do exist a boundary between the two phases. However, configurations in percolation model are obtained randomly without any dynamical constrains, which leads to a non-clustering figure in the PCA's diagram (Fig. 3 (c)). Hence, the K-means algorithm gives a trivial classification, i.e., it equally divides the data into two parts. Meanwhile, the confusion scheme introduces a high degree of confusion at $c$ because the non-percolating and percolating configurations are mixed around this point (Fig. 3 (d)). Therefore, the unsupervised learning can be applied only for models which configuration vectors can be clustered by PCA.

\begin{figure}
  \centering
  \includegraphics[width=0.5\textwidth]{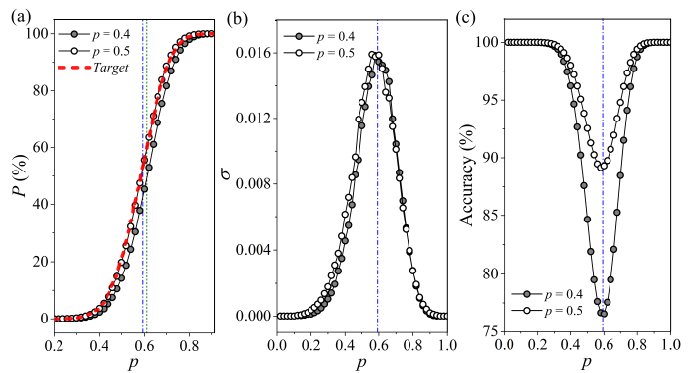}\\
  \caption{Results of supervised learning of the percolation model. (a), (b) and (c) respectively denotes the predicted percentage $P$, standard deviation and accuracy versus open probability $p$. The filled and open circles give the results of the learning machine trained by samples at $p=0.4$ and $p=0.5$, respectively. Vertical dashed lines denote the critical point of the percolation model and the red line in (a) is the standard results by the conventional approach. The solid lines in the three figures are drawn for guiding the eyes.}\label{4}
\end{figure}

\emph{Supervised Learning --- }Firstly, for the percolation model, we construct two sets of $10^4$ labeled configurations at two points $p=0.4$ and $p=0.5$ far from the critical point $p_c \approx 0.593$, and use them to train two CNNs. In Fig. 4 (a), we plot the predicted phase transition curves of the two trained CNNs from the $n=10^5$ unlabeled configurations in the interval $p \in [0, 1]$. For comparison, the target curve is provided by the two-pass algorithm \cite{shapiro2001computer}. It can be seen that the predicted curves are very close to the target one. Especially, that of $p=0.5$ is visually indistinguishable from the target. An analysis indicates that the $10^4$ training samples obtained at $p = 0.4$ include only $403$ percolating configurations, which is about $0.4\%$ of the total number of samples. This is quite a low proportion. Even the training samples obtained at $p = 0.5$ include only approximately $20\%$ percolating configurations. Note that the geometries of the percolation configurations at different value of $p$, particularly those at low and high $p$, have remarkable difference \cite{SupplementMaterial}. Hence, the generalization ability of the learning machine is quite surprising, which can extend limited knowledge it learned to more general situations. This is the basis that the supervised learning can be applied to study the phase transition.

\begin{figure}[t]
  \centering
  \includegraphics[width=0.5\textwidth]{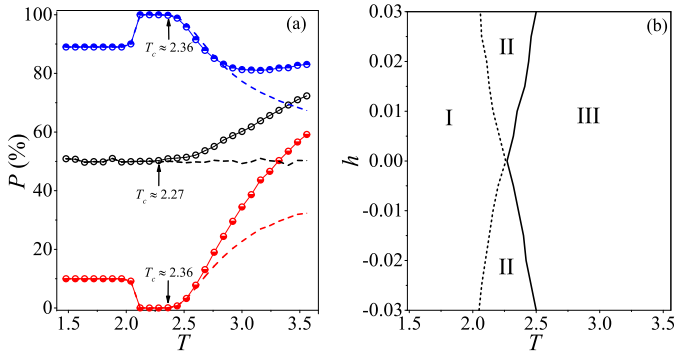}\\
  \caption{Results of supervised learning of Ising model. (a) The proportion $P$ of up-dominant configurations versus temperature $T$ predicted by the learning machine (circles) and by a direct counting process (dashed lines) for external field $h$ values of $-0.02$, $0$, and $0.02$ (from bottom to top). (b) The phase diagram predicted by the learning machine, which is divided into a low-temperature ferromagnetic phase region (I), a high-temperature ferromagnetic phase region (II), and a non-ferromagnetic phase region (III).}\label{5}
\end{figure}

The learning machine of $p=0.4$ shows a slight deviation from the target. Indeed, one can find out that the critical point obtained by this learning machine is about $p_c \approx 0.61$. As such, the accuracy of the learning machine may affect the accuracy for identifying the critical point when applying the traditional method of learning machine to study the phase transition. Thus we seek for an accuracy-independent way to identify the critical point. We divide the testing set into $100$ subsets at each point of $p$, and calculated the standard deviations $\sigma$ of the predicted proportion for each $p$ when using the training data sets obtained at $p=0.4$ and $p=0.5$, respectively, as shown in Fig. 4 (b). The significant feature here is that the critical point appears at the maximum values of $\sigma$ for both CNNs. Moreover, these two curves has almost no obvious difference on the critical point identification. To reveal the underlying mechanism, we show accuracies of these CNNs on the testing sets along $p$, see Fig. 4 (c). It can be seen that there is a minimum on the accuracy curve, where the position of it exactly corresponds to the critical point, and is independent with the accuracy. This fact indicates that the information of the critical point can be incorporated into the learning machine, even when trained by the data set obtained far from the critical point. The biggest uncertainty of $\sigma$ at the critical point is only induced by the worst performance. This finding inspires a accuracy-insensitive strategy to determine the critical point: regardless of the accuracy, searching for the singularity in the accuracy can always find the correct critical point.

This strategy can also be applied to the Ising model. We evolve the system a sufficiently long period at the parameter point $(h,T) = (0,3.5)$, which is far from the critical point, to obtain $10^4$ configurations. They are classified into two classes by labeling a configuration by the sign $S$ which can be defined as $\sum_{i=1}^N \sigma_i$. The predicted percentages $P$ of up-dominant configurations ($S=1$) are presented by circles in Fig. 5 (a) as a function of $T$ for $h= -0.2$, $0$, and $0.2$, respectively. For comparison, the dashed lines represent the accurate percentages obtained by direct counting.

We find that for each $h$, the predicted curve of the proportion $P$ deviates from the target one in the high-temperature region, indicating that the accuracy is indeed low there. With the decrease of the temperature, the prediction approaches to the result of direct counting, and the deviation vanishes under the critical point. This phenomenon can be recovered if we apply another training set. We find that the critical points appear as the turning points that the gradient of the predicted curves turn to vanish, where the values of them are assigned by arrows in Fig. 5 (a). Therefore, the learning machine, independent of its accuracy on the high-temperature region where it is trained, can identify the critical point. In such a way, it can automated recover the phase diagram of the target system, see Fig. 5 (b).

The underling mechanism is explained as follows. In the non-ferromagnetic phase, $S$ takes the Gaussian distribution. The up-dominant and down-dominant classes are divided when $S = 0$. If the boundary of a learning machine to classify them is perfectly consist to the standard boundary, all of the configurations can be correctly classified. However, the boundary of classification often has a bias to the standard one when trained by a finite training set, and it is independent of the training algorithm. The bias, even very slight, may result a remarkable error rate when the Gaussian distribution lies around the region of $S = 0$. Note that the error is proportional to the area between the boundary of classification and the standard one. This situation occurs in the high-temperature region. With the decrease of the temperature, the peak of the distribution shifts, so that the standard boundary lies across the tail part of the Gaussian distribution. Since the tail of the Gaussian distribution decays in the manner fast than the exponential, the error of the learning machine decreases very quickly. Below the critical point, the Gaussian distribution function shift above or below completely to the line of zero, resulting a gap there. Only when the boundary of classification of a learning machine locates in the gap, the error rate vanishes \cite{SupplementMaterial}.

Here a relatively inaccurate learning machine may be helpful. As a particular example, in the case of $h = 0$, the direct counting should give a constant curve of $P=0.5$ independent of the temperature because of the symmetry of the system, which provides no information of the critical point. However, the learning machine in Fig. 5 (a) still reveals the critical point following the criterion that the gradient turns to vanish.

The plateaus of $P=1$ for $h>0$ and $P=0$ for $h<0$ represent the symmetry breaking that each configuration takes a definite positive or negative magnetization slightly under the critical point. Another type of plateaus left to plateaus of $P=1$ and $P=0$ indeed reveal a sub-phase. In this phase, certain initial configurations may evolve to a stationary state with positive magnetization for $h>0$. Similar phenomenon appears when $h<0$. Particularly, it is interesting that the proportion of these initial configurations keep to be constant in this phase, independent of the temperature. Further studies show that plateaus in this phase are finite-size effects, since they may eventually approach the plateaus of $P=1$ or $P=0$ with the increase of the system size. Nevertheless, it can also be an inevitable effect in nanoscale systems. To the best of our knowledge, this phenomenon has not been documented in Ising-like models. If concatenate the turning points on the two edges of the plateaus of $P=1$ and $P=0$, two lines divide the parameter plain $(h,T)$ into three regions: a low-temperature ferromagnetic phase region (I), a high-temperature ferromagnetic phase region (II), and a non-ferromagnetic phase region (III) shown in Fig. 5 (b).

\emph{Summary --- }One can apply the learning machine in a non-traditional way to design automated algorithm to study phase transitions. With the new strategy, the purpose of training is to establish the intrinsic correlation between the learning machine and the target system, and only a training set made at a non-critical parameter is fascinatingly sufficient to fulfil this purpose. The critical point, as well as other possible inter-phase transition points, can be identified according to the global behavior of the learning machine. This method is insensitive to the accuracy of the learning machine. We hope this strategy open a new road to study physical problems using learning machine.

We acknowledge the support by NSFC (Grant No. 11335006).

\bibliography{References}

\end{document}